\documentclass{article}
\usepackage{amsfonts}
\usepackage{amssymb}
\usepackage{amsmath}
\usepackage{amsthm}
\usepackage[english]{babel}
\usepackage{epigraph}
\usepackage{graphicx}

\begin{document}

\title{\bf A Pamphlet against The Energy}

\author{{\bf Alexey Golovnev}\\
{\small {\it Centre for Theoretical Physics, The British University in Egypt,}}\\
{\small \it El Sherouk City, Cairo 11837, Egypt}\\
{\small agolovnev@yandex.ru}}
\date{}

\maketitle

\epigraph{What would the people, so empty and limited, say \\
If just out of fantasy, on a randomly chosen day,\\
Silver-violet shades had I dyed all my hair,\\
In an ancient Greek gown, was my comb's betrayer:\\
Myosotis or jasmines on my head all around,\\
And I sang in the streets to the violins' sound,\\
Or presented my verses in the squares and aisles\\
With the freedom of taste vagabonding the miles?\\
Would they stare and look for a better position?\\
Would they burn me, a witch for a new inquisition?\\
Would they ring all the bells calling masses to pray?\\
Such a wonderful image that has made my day.}{Alfonsina Storni. ?`Qu{\'e} dir{\'i}a?\\
(My translation)}

The problem of energy, or more generally of all the standard conservation laws related to the Poincar{\' e} symmetry group, is known since the very first years of modern gravity theories. And an immediate reaction which should come to our mind when seeing the common panic about the non-existent notion of energy is "WHY?". Why is it taken as a problem, to start with? Why don't we just accept that there is no meaningful notion of conserved energy, and that's it?

We are so much used to having a well-defined energy in all our classical physical theories that we tend to fully forget that the very notion is quite artificial. It's never something we measure directly, if in absence of fundamental gravity. Instead, it's a conserved quantity which follows from symmetry under time translations. If we measure the velocity of a body of a known mass, we can calculate its kinetic energy, thus doing an indirect measurement, then after a crash it goes into heating up the environment, and we can check the energy conservation by now measuring the temperature, one way or another.

There is no natural device for which the energy itself could be taken as a directly measurable quantity. However, it is a very convenient tool which very often incredibly helps us solve equations of motion or prove some general properties of solutions, let alone its instrumental role in statistical mechanics. Locally, and when without really strong gravity, one can introduce the Newtonian potentials and still have the advantage of conservation laws, even in presence of gravitational forces.

Note though that, even in the most classical physics, the conserved quantities do depend on the chosen reference frames. Getting some simple conservation laws, with no explicit time variable in the definitions, requires what is called inertial frames, and even there the energy and momentum are not uniquely defined. In relativistic theories, we combine them into an energy-momentum 4-vector which is finally a quantity independent of the choice of a frame, thus reflecting an observed symmetry of the world at this level. Though the very definition still has an ambiguity in it. One can multiply a definition of a conserved quantity by a universal constant, say $\frac{e^{\pi^2}}{\sqrt{17}}$, and it will remain a conserved quantity.

The situation becomes even more complicated in the field theory where we get to use the energy-momentum tensor. Then, in an arbitrary pseudo-Riemannian spacetime, the old usual conservation law $\partial_{\mu} T^{\mu\nu}=0$ transforms into the covariant "conservation" $\bigtriangledown_{\mu} T^{\mu\nu}=0$, with the Levi-Civita covariant derivative. It does not correspond to an actual conservation of any integral quantity. It rather says that $\frac{1}{\sqrt{-g}}\partial_{\mu}\left(\sqrt{-g} T^{\mu\nu}\right) + \Gamma^{\nu}_{\mu\alpha} T^{\mu\alpha}=0$ which, due to the last term, is not precisely a conservation. And, of course, if the spacetime had some particular symmetry, then there might have been some symmetry-preferred coordinates in which the unwanted term is killed, or even better a contraction of $T^{\mu\nu}$ with the Killing vector, and we get a well-defined conservation law as long as the symmetry is protected.

Note that the vectorial things like electric charge conservation $\bigtriangledown_{\mu} J^{\mu}=0$ do have a meaningful integral conservation law via $\bigtriangledown_{\mu} J^{\mu}=\frac{1}{\sqrt{-g}} \partial_{\mu} \left(\sqrt{-g} J^{\mu}\right)$, while generically it is not the case for the energy behaviour in gravity. This is just a mere fact to accept: what was described as $\partial_{\mu} T^{\mu\nu}=0$ in the Cartesian coordinates of a flat space cannot correspond to a covariant conserved quantity in general. At the same time, the covariant "conservation" is still a good enough relation to use. For example, it is very convenient and important to know how the energy density $\rho$ of cosmological matter evolves: $\dot\rho + 3H(\rho + p)=0$ in the "physical" time (the one with unit lapse) and with the Hubble "constant" $H$, and $p$ being the pressure. It is not about any conservation, but it gives us all the information we need.

\begin{figure}[t!]
\includegraphics[width=\linewidth]{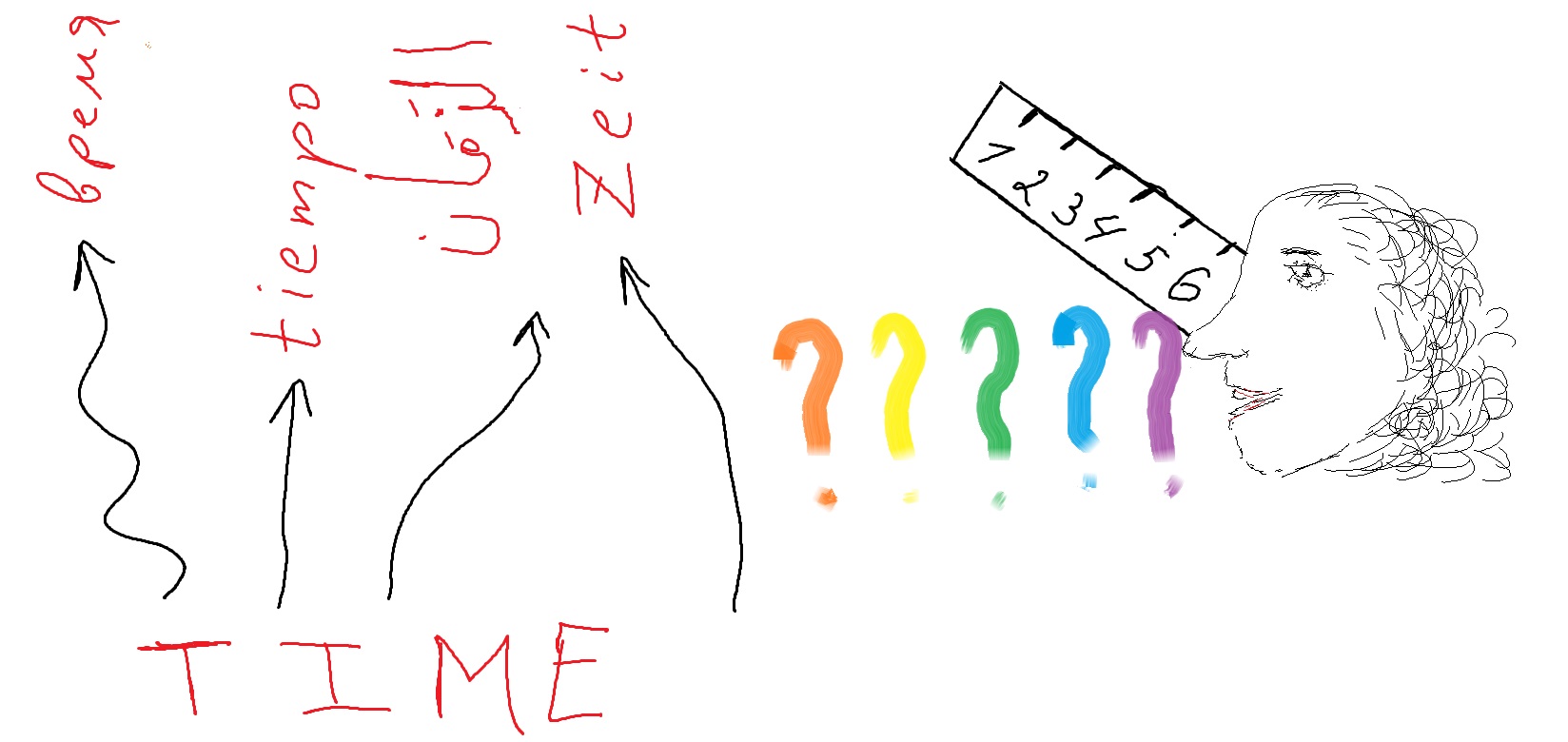}
\end{figure}

Of course, many people have been able to find lots of conserved "pseudo-tensors", which I would rather call quasi-tensors, since the prefix "pseudo" is often used for special relativistic almost tensors, the ones which violate the symmetry with respect to only the disconnected components of the Lorentz group. And it makes perfect sense. Using particular coordinate systems, one can find conserved quantities indeed, however there is no covariant notion of this conservation simply because there is no well-defined and covariant symmetry behind it.

There is no problem of defining {\it some} conservation, and we can even do it in a more beautiful way. For example, if the spacetime has trivial topology, or we are interested in only a single coordinate chart on it, then what we have got is topologically nothing else but a Minkowski space. This fake Minkowski has all the needed symmetries and one can have a perfectly well-defined energy-momentum tensor on it. The only problem is that generically it would be something esoteric for a real observer. 

Following this line, if gravity is not too strong, one can take the metric as $g_{\mu\nu}(x)=\eta_{\mu\nu} + h_{\mu\nu}(x)$, or similar,  and we get it as a special relativistic field theory of the $h_{\mu\nu}(x)$ field in the fictitious Minkowski spacetime. One can then define the energy for gravitational waves, and construct the whole theory {\` a} la Feynman \cite{Feyn}. This Minkowski space is something voluntaristic put upon the theory by hand. However, it still allows one to do pretty much, even totally disregarding all the fundamental geometry behind, as for example Weinberg liked to do \cite{Wein}.

I don't like ignoring the geometric structure of General Relativity even though, technically, it is a perfectly legitimate approach. We simply make the actual spacetime correspond to some abstract Minkowski space, of course in absolutely not unique a way. However, going this path too far and with certain lack of common sense, one can end up constructing the so-called relativistic theory of gravity \cite{log}. It would even be correct if they hadn't claimed difference from the standard GR.

Yet another option is to walk in the shoes of Regge and Teitelboim \cite{RT}, with our 4D spacetime being a submanifold in a 10D Minkowski spacetime. Then, in the flat ambient space, we do have a well-defined conserved energy-momentum tensor for the full theory, though no less esoteric for the mundane creatures than the one defined around an arbitrarily introduced 4D Minkowski. Actually, this theory generalises the General Relativity way too much, and therefore has largely lost all its one-time popularity, though some funny games are definitely possible with it \cite{PaFr}.

All in all, one can find lots of versions of conservation laws, and this is perfect. However, it has no relation to a covariant conserved quantity, for that simply does not exist. The deep reason behind is that we have a full freedom of coordinate choice, and then it's not even clear what is time with respect to which the conservation is expected, nor what is space to measure the volumes. On the other hand, in a generic spacetime, there is no covariant realisation of the symmetry either, the one which would be related to the desired conservation, as well as no natural notion of inertial frames. As a substitute, and in a fully arbitrary manner, we introduce a fictional Minkowski space, either as a higher dimensional realm full of ghosts or as a projection to remote invisible heavens, and then we do have a perfectly covariant but pretty esoteric mantra of conservation laws.

\begin{figure}[t!]
\includegraphics[width=\linewidth]{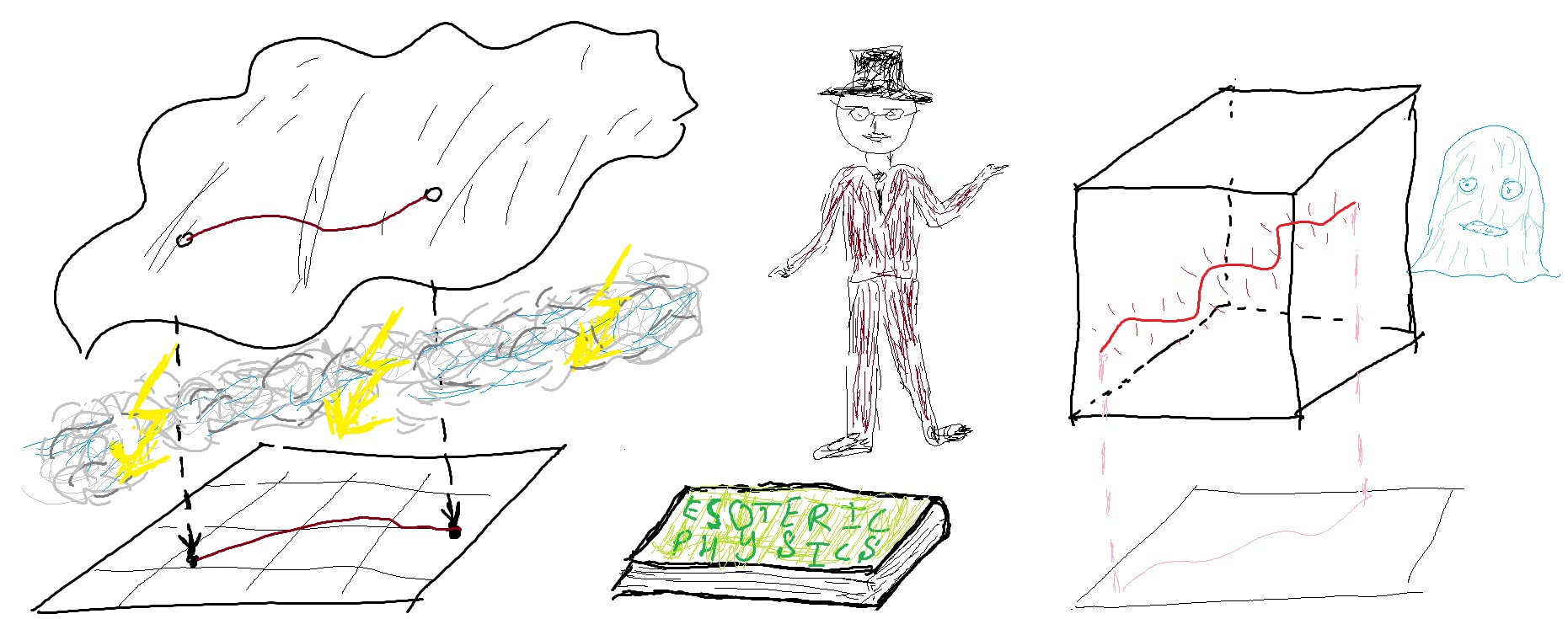}
\end{figure}

At the same time, in case of a symmetry available, there is a preferred choice of coordinates and one can meaningfully speak of some down-to-earth and physical conservation. In particular, if we assume the spacetime be asymptotically flat, then the commonly preferred, and geometrically natural choice is to have coordinates tending to the Cartesian ones at infinity, and we end up with well-defined global conserved quantities. However, having made a step towards the so-called "quasi-local" ones, it's immediately possible to define many different versions of those \cite{ql}, potentially igniting heated discussions on which ones would be our favourites. I avoid using the adjectives of "right" or "wrong" for discussing the choice of an objectively meaningless quantity.

Then, one more option came from the pure-tetrad description of teleparallel gravity \cite{Moller, PereiraM}. Oversimplifying the whole story, if we assume a given physical tetrad $e^{\mu}_a$ and find a physical quantity ${\tilde T}^{\mu}_a$, or $J^{\mu}_a$, with the mixed indices, such that $\partial_{\mu}\left(\sqrt{-g} {\tilde T}^{\mu}_a\right)=0$, then it is a real conservation. It was discarded though \cite{PereiraM} for being dependent on the choice of the tetrad.

I would say that this particular objection comes from misunderstanding the geometry of teleparallelism. It should not be locally Lorentz invariant \cite{meW}. The common perception some people have \cite{PereiraM} is that the tetrad is just a soldering form, and mustn't be anything except rewriting the vectors and other tensors in an arbitrary tangent space basis. However, it is not the case of teleparallel gravity, especially not when modified.

In particular, the zero spin connection has a very special meaning in teleparallel theories \cite{meW}. Those tetrads are bases of vector fields which are covariantly constant: $\bigtriangledown_{\mu} e^{\nu}_a =\partial _{\mu} e^{\nu}_a + \Gamma^{\nu}_{\mu\alpha} e^{\alpha}_a =  0$ with a teleparallel connection $\Gamma$. Therefore, a fundamental tetrad should not then be thought of as a free choice of an observer. Quite to the opposite, it must be taken as the defining notion of the geometry. The teleparallel tetrad is a set of covariantly constant vectors, and it is neither a rank two tensor nor a mere soldering form. It is quite similar to the case of Cartesian coordinates. A Euclidean space can be perfectly written down in terms of arbitrary curvilinear coordinates, however the Cartesian ones are its objectively defining features.

In this case, the conservation law of  the $\partial_{\mu}\left(\sqrt{-g} {\tilde T}^{\mu}_a\right)=0$ shape can be taken as ${\mathop\bigtriangledown\limits^{(0)}}_{\mu} {\tilde T}^{\mu}_{a}=0$, where by the superscript $\mathop{}\limits^{(0)}$ in the teleparallel context we mean the quantities related to the Levi-Civita connection, and therefore it is a perfectly covariant  law which states that Levi-Civita divergences  of the four different vectors ${\tilde T}^{\mu}_a$ vanish. Such vectors can be constructed as projections of a rank two tensor onto the basis of objectively preferred vectors $e^{\mu}_a$. Thinking of the tetrad as a soldering form won't help anything in these terms, for it would be just a rewriting of a symmetric tensor covariant divergence which cannot be interpreted as an integral conservation law. However, when treated as four separate vectors, their conservation does have a covariant meaning. But for us, the mundane researchers, it is again some artificially imposed curvatureless parallel transport structure, though with its notion of conservation.

Let me explain the meaning of teleparallel geometry once more \cite{meW}. We assume that the spacetime has got a connection of vanishing curvature on it. It means that, having chosen a basis of tangent vectors at some point, we can parallelly transport it to every other point of the manifold and get a uniquely defined field of the tangent space bases. Of course, there is a global Lorentz freedom of choosing it, however otherwise it is unique, thanks to the absence of holonomy in this version of parallel transport. It provides us with a proper substitute for the needed symmetry, or with a new notion of inertial frames.

A few years ago, another variant of teleparallel gravity, symmetric teleparallel, became popular \cite{coin}. It is basically the same idea as with the metric teleparallel, though it works in terms of non-metricity instead of torsion. In practical terms, it means that there exist some particular coordinates $\xi^{\mu}$ in which the affine connection coefficients are simply zero. In other words, it is a bona fide hidden Minkowski. More precisely, there exist coordinates $\xi^{\mu}$ in terms of which the parallel transport is that of a Minkowski spacetime with $\xi^{\mu}$ being its Cartesian coordinates, and with their coordinate basis in the role of the fundamental teleparallel tetrad \cite{BGGM}.

Having fixed the coordinates to the ones with zero affine connection coefficients, we are in what is called the coincident gauge; and the symmetric teleparallel equivalent of General Relativity can be written as the historic $\Gamma\Gamma$ action of Einstein. Unlike the case of the proper action due to Hilbert, its Lagrangian density is not a scalar. And then comes the canonical energy-momentum "tensor" proposed by Einstein as a measure of the fictitious gravitational energy and momentum. Of course, it badly depends on the choice of coordinates.

In a sense, it takes us back to the field-theoretic view on gravity. We treat the Riemannian metric as a field on a rather trivial manifold, the one which has got the same parallel transport as the good old Minkowski spacetime had, and fix the coordinates $\xi^{\mu}$ to serve the same role as did the Cartesian ones in the special relativistic context. In sufficiently modified teleparallel theories, these structures might get enough of objectively dynamical meaning. However, in GR-equivalent models, they are nothing but an ideal paradise we have imagined out of nothing.

Recently, there was a progress in this respect, namely that of covariantisation \cite{JT1, JT2}. To the energy problem though, it doesn't do anything more than just rewriting the Einstein's quasi-tensor in a formally covariant form. There were coincident gauge coordinates $\xi^{\mu}$ with zero connection, and with the gravitational "energy" proposal by Einstein. Now, precisely the same thing gets rewritten in other coordinates. The only difference \cite{BGGM} is that we have allowed ourselves to use whatever coordinates $x^{\mu}$ we like, and promoted the previously preferred coordinates to a set of scalar fields $\xi^{\mu} (x)$.

\begin{figure}[t!]
\includegraphics[width=\linewidth]{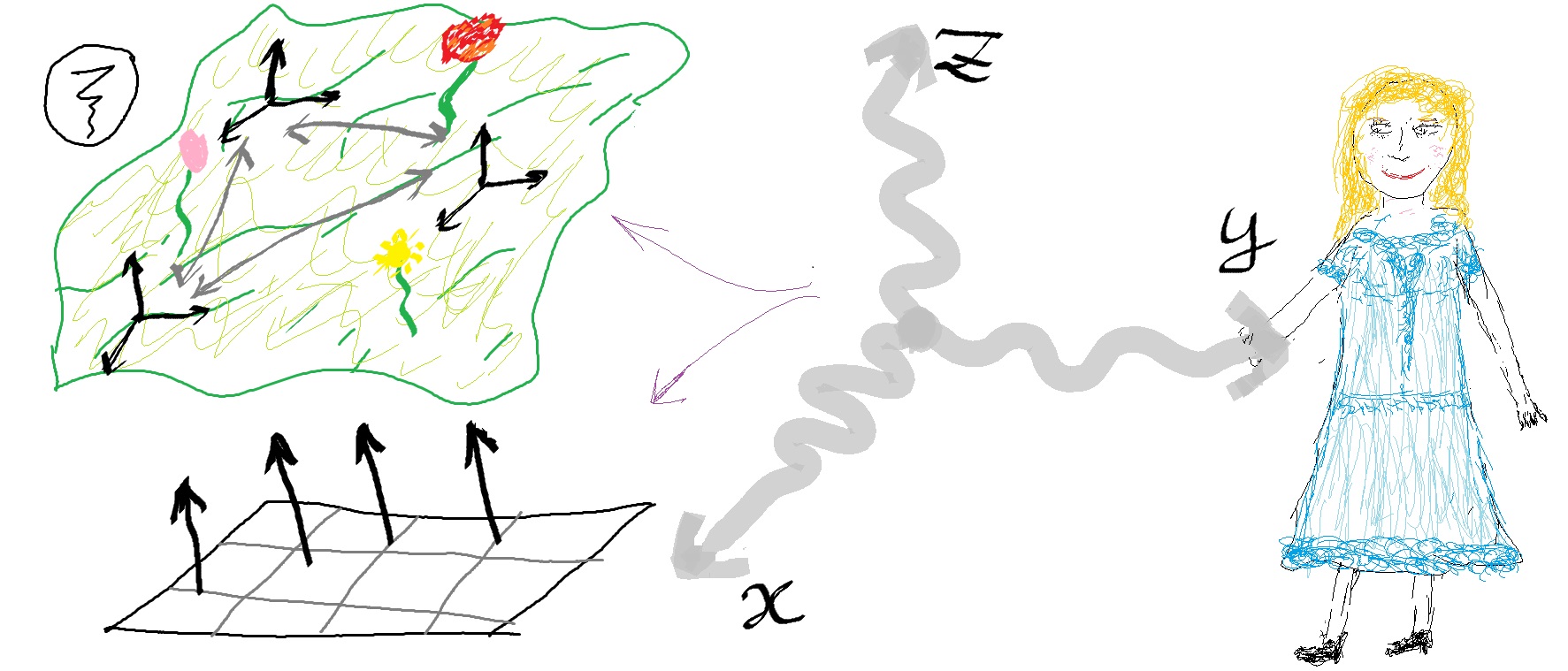}
\end{figure}

Summarising the symmetric teleparallel approach, it is a genuine quintessence of the field-theoretic viewpoint. A trivial Minkowski spacetime lies in the base, and in the coincident gauge we simply take its Cartesian coordinates as the ones which have all the connection coefficients vanishing. The model has then got conserved quantities, related to the unobservable Minkowskian structure. The really observable Riemannian geometry is treated as just a dynamical field on top of the esoteric thing with its own conservation laws. The covariant formulation simply rewrites it in arbitrary coordinates. We trade the preferred coordinates for fixed functions on the manifold, the same way as the standard Cartesian coordinates can be treated as some very special scalar functions on a Euclidean space.

My main message is that all our attempts at defining a covariant conserved energy are very artificial constructions with no good meaning behind. Why are we doing it then? Let's not discuss Black Hole thermodynamics. It is an extremely interesting outcome in itself  that thermodynamics does seem to work when there is no good reason for why it should, if not to consult the far-fetched constructions of string theory. Probably, the most important motivation for the unnatural quantities comes from the wish to quantise gravity. There is indeed a big problem of making the gravitational physics and quantum physics compatible with each other. However, the next question to ask is whether it is really the case that what should be changed is our understanding of gravity, and not the quantum.

\begin{figure}[t!]
\includegraphics[width=\linewidth]{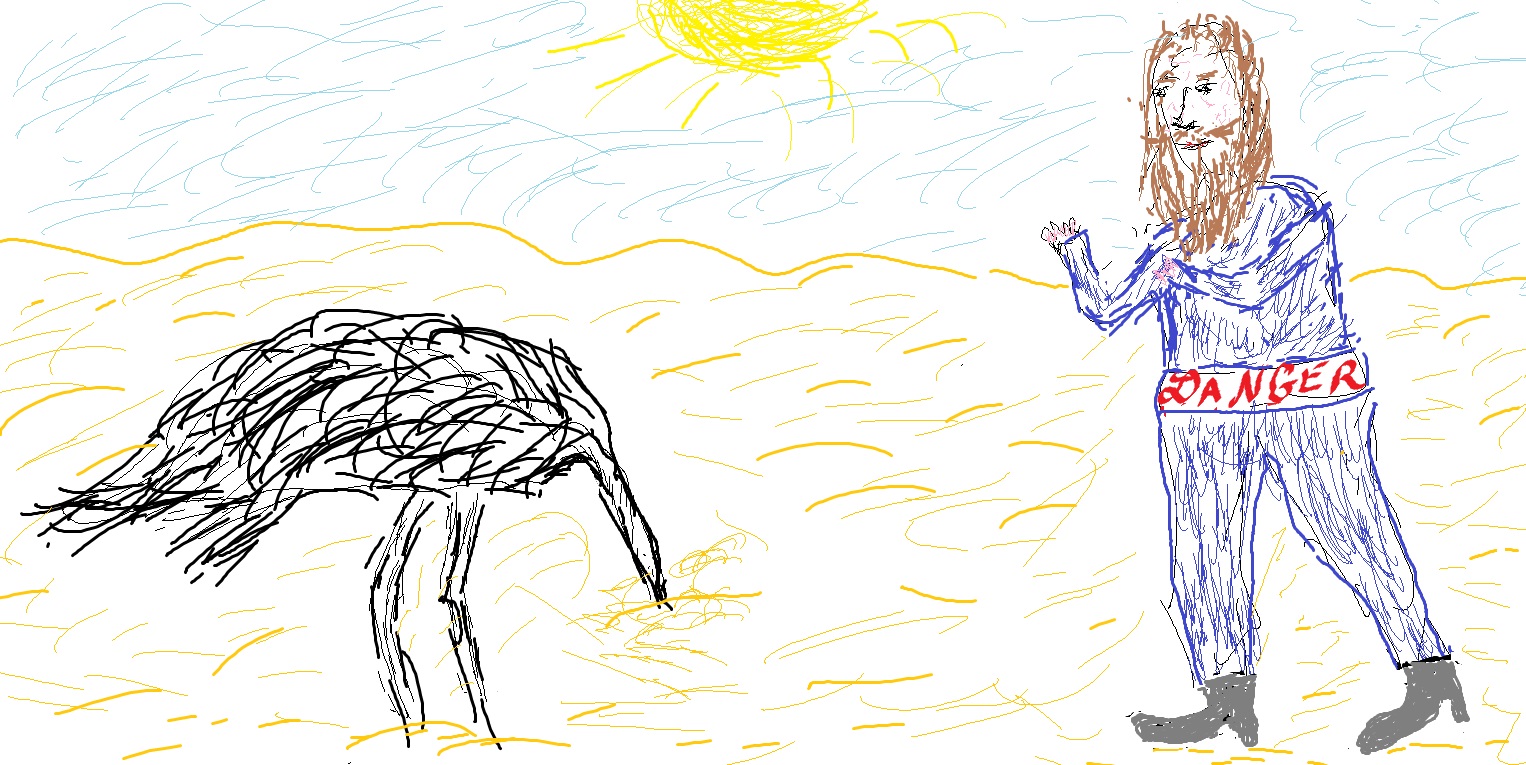}
\end{figure}

There are many facets of this issue, from technical problems of non-renormalisability to the foundational questions of time. The latter point is very important indeed, and to some extent we have discussed it above. As to the former, and intentionally exaggerating everything, I would formulate my attitude to the formal parts as the claim that {\it quantum physics simply makes no good sense at all, except for the mere fact that it fantastically corresponds to all experimental data}. 

At the level of non-trivial quantum field theory we just lack any good understanding of the mathematical structure behind \cite{Ded}, even though there are good reasons to believe that our general approach is how it should be \cite{WeinQ}. And on top of all that, even for the basic quantum mechanics \cite{WeinQQ}, there is the infamous measurement and interpretation problem. Maybe, what we have to change is not gravity.

It seems that, once we've faced a problem, we hide our head in the sand. If we avoid seeing the culprit, it's probably not really here. The instantaneous wave function collapse cannot lead to troubles, due to linearity of quantum mechanics, let's then shut up and calculate. If quantum physicists need a notion of conserved energy, let's then pretend it is there, even in the gravitational worlds. It is a common type of reaction, I must admit. Let me remind you though that literally digging the head into the sand is just a myth, even for the ostrich. Moreover, upon a bit of serene thinking, the absence of energy does not seem more dangerous than the guy I've sketched in the picture.

In somewhat the opposite direction, there is also an interesting cultural aspect. We've seen something unusual, that is absence of a well-defined energy. And then we immediately start asking questions in the path of Russian classical literature: "Who is to blame?" by Alexander Herzen and "What is to be done?" by Nicolay Chernyshevsky. I am judging neither the love affairs of these novels nor their political or philosophical lines. The funny point is that these are commonly taken as the ageless Russian questions. However, it seems that many international research reactions aren't very far from that. 

And in our case we might simply need to realise that the problem we are discussing is just an imaginary one. If a successful theory does not offer us something we would like to have, it's probably not a good reason to force the theory, it's time to rethink our wishes. Most importantly, General Relativity has no preferred frames, by its very geometric nature. Unless we modify the theory very substantially, we have to accept it.

I do agree that it potentially speaks of a very serious contradiction to our current understanding of quantum physics, such as the Unruh effect \cite{Unruh} and around. One way to go is to try to change the quantum realms, and this is what I would vote for; another way is to have a background spacetime for the quantum vacuum, with one of the cleanest examples in the "canonical frame" \cite{JT1} of symmetric teleparallelism. The latter is definitely possible but absolutely alien to the classical gravity we know.

Coming back to the epigraph, I conclude. Barring possibly very big changes, it seems that the known gravitational physics is telling us to forget the notion of the fundamentally conserved energy and to not hope for a general luxury of inertial frames. Of course, for some people it might look indeed like throwing a comb away (though a bit old-fashioned by now, a big decorative comb to be worn under a mantilla), and they might even take it as going savagely wild. However, there is nothing divine about traditions, and we must always be ready for reassessing all what we are used to.


\begin{thebibliography}{99}

\bibitem{Feyn} R.P. Feynman, F.B. Morinigo, W.G. Wagner, B. Hatfield. Feynman Lectures on Gravitation. Addison-Wesley, Boston 1995.

\bibitem{Wein} S. Weinberg. Gravitation and Cosmology: Principles and Applications of the General Theory of Relativity. Johm Wiley \& Sons 1972.

\bibitem{log} A.A. Logunov, Yu.M. Loskutov,  M.A. Mestvirishvili. {\it Relativistic theory of gravitation and its consequences}. Progress of Theoretical Physics, {\bf 80} (1988) 1005

\bibitem{RT} T. Regge, C. Teitelboim. {\it General Relativity {\` a} la string: a progress report}. Proceedings of the First Marcel Grossmann Meeting, Trieste, Italy, 1975. North-Holland, Amsterdam, 1977. arXiv:1612.05256

\bibitem{PaFr} S.A. Paston, V.A. Franke. {\it Canonical formulation of the embedded theory of gravity equivalent to Einstein's General Relativity}. Theoretical and Mathematical Physics {\bf 153} (2007) 1581; arXiv:0711.0576

\bibitem{ql} L.B. Szabados.{\it Quasi-Local energy-momentum and angular momentum in General Relativity}. Living Reviews in Relativity {\bf 12} (2009) 4

\bibitem{Moller} C. M{\o}ller. {\it Conservation laws and absolute parallelism in General Relativity}. Matematisk-fysiske Skrifter udgivet af Det Kongelige Danske Videnskabernes Selskab Bind 1, nr. 10. K{\o}benhavn 1961

\bibitem{PereiraM} V.C. de Andrade, L.C.T. Guillen, J.G. Pereira. {\it Gravitational energy-momentum density in teleparallel gravity}. Physical Review Letters {\bf 84} (2000) 4533; arXiv:gr-qc/0003100

\bibitem{meW} A. Golovnev. {\it The geometrical meaning of the Weitzenb{\" o}ck connection}. International Journal of Geometric Methods in Modern Physics; arXiv:2302.13599

\bibitem{coin} J. Beltran Jimenez, L. Heisenberg, T.S. Koivisto. {\it Coincident General Relativity}. Physical Review D {\bf 98} (2018) 044048; arXiv:1710.03116

\bibitem{BGGM} D. Blixt, A. Golovnev, M.J. Guzman, R. Maksyutov. {\it Geometry and covariance of symmetric teleparallel approaches to gravity}. arXiv:2306.09289

\bibitem{JT1} J. Beltran Jimenez, L. Heisenberg, T.S. Koivisto. {\it The canonical frame of purified gravity}. International Journal of Modern Physics {\bf 28} (2019) 1944012; arXiv:1903.12072

\bibitem{JT2} D. Aguiar Gomes, J. Beltran Jimenez, T.S. Koivisto. {\it Energy and entropy in the geometrical trinity of gravity}. Physical Review D {\bf 107} (2023) 024044; arXiv:2205.09716

\bibitem{Ded} M. Dedushenko. {\it Snowmass White Paper: The Quest to Define QFT}. International Journal of Modern Physics A {\bf 38} (2023) 2330002; arXiv:2203.08053 

\bibitem{WeinQ} S. Weinberg. The Quantum Theory of Fields I. Cambridge University Press 1995.

\bibitem{WeinQQ} S. Weinberg. Lectures on Quantum Mechanics. Cambridge University Press 2015.

\bibitem{Unruh} W.G. Unruh. {\it Notes on black-hole evaporation}. Physical Review D {\bf 14} (1076) 870

\end{thebibliography}
\end{document}